\documentclass[prb,aps,floatfix,onecolumn,superscriptaddress,amsmath]{revtex4}

\usepackage[caption=true,subrefformat=parens,labelformat=parens]{subfig}
\usepackage{graphicx}
\usepackage{float}
\usepackage{siunitx}
\usepackage{amsmath}
\usepackage{bm}
\usepackage{multirow}
\usepackage{xcolor}
\usepackage{xfrac}
\usepackage{chemformula}

\newcommand{\be}{\begin{equation}}
\newcommand{\ee}{\end{equation}}
\newcommand{\bea}{\begin{eqnarray}}
\newcommand{\eea}{\end{eqnarray}}

\def\nn{\nonumber}
\def\lb{\label}

\def\bk{{\bf k}}

\def\bq{{\bf q}}

\def\bQ{{\bf Q}}

\def\d{\delta}

\def\t{\tau}

\def\etap{\eta^{\prime}}
\def\deltap{\delta^{\prime}}
\def\s{\sigma}

\def\G{\Gamma}

\begin{document}
\title{The role of orbital nesting in the superconductivity of Iron-based Superconductors}

\author{Raquel Fern\'andez-Mart\'in}
\affiliation{Materials Science Factory, Instituto de Ciencia de Materiales de Madrid, ICMM-CSIC, Cantoblanco, 28049 Madrid, Spain}

\author{, Mar\'ia J. Calder\'on}
\affiliation{Materials Science Factory, Instituto de Ciencia de Materiales de Madrid, ICMM-CSIC, Cantoblanco, 28049 Madrid, Spain}

\author{Laura Fanfarillo}
\affiliation{Scuola Internazionale Superiore di Studi Avanzati (SISSA), Via Bonomea 265, 34136 Trieste, Italy}
\affiliation{Department of Physics, University of Florida, Gainesville, Florida, USA}

\author{Bel\'en Valenzuela}
\affiliation{Materials Science Factory, Instituto de Ciencia de Materiales de Madrid, ICMM-CSIC, Cantoblanco, 28049 Madrid, Spain}

% abstract 
\begin{abstract}{We analyze the magnetic excitations and the spin-mediated superconductivity in iron-based superconductors within a low-energy model that operates in the band basis but fully incorporates the orbital character of the spin excitations. We show how the orbital selectivity, encoded in our low-energy description, simplifies substantially the analysis and allows for analytical treatments, while retaining all the main features of both spin-excitations and gap functions computed using multiorbital models. Importantly, our analysis unveils the orbital matching between the hole and electron pockets as the key parameter to determine the momentum-dependence and the hierarchy of the superconducting gaps, instead of the Fermi surface matching as in the common nesting scenario.}
\end{abstract}

\maketitle

The discovery of iron-based superconductors (IBS) raised immediate questions about the nature of the superconducting (SC) state and the pairing mechanism. From the very beginning it was proposed that pairing could be unconventional \cite{Mazin2008, Kuroki2008}. This proposal has been triggered by both the small estimated value of the electron-phonon coupling \cite{IBSnophonon} and the proximity in the phase diagram of a magnetic instability nearby the SC one. Within a band-nesting scenario, pairing is provided by repulsive spin-fluctuations between hole and electron pockets, connected by the same wave vector characteristic of the spin modulations in the magnetic phase \cite{Chubukov_Chapter2015}. 
Given the repulsive and interband character of the interaction, the expected symmetry for the gap function is the so-called $s_{\pm}$, i.e. an isotropic s-wave on each pocket with opposite signs for hole and electron pockets.
This picture has been supported and confirmed by extensive theoretical works that, within realistic multiorbital interacting models for IBS, provide a quantitative estimate of the SC properties starting from a Random Phase Approximation (RPA)-based description of the spin-susceptibility \cite{Graser2009, Arita2010, Graser2010, Acommonthread, GapsymmetryHirschfeld2011, GapanisotropycomparitionRPA, ReviewHirschfeld}. 

The inclusion of the orbital degree of freedom in the analysis of the SC gaps allows to reproduce a number of features experimentally observed in IBS \cite{ARPESreview} such as the angular modulation of the $s_{\pm}$ gap functions and the possibility of accidental nodes on the Fermi surfaces. 
On the other hand, the number of orbitals included makes the analysis of superconductivity within multiorbital models very complex. As a consequence analytical treatments of the problems are often unattainable and the physical interpretation of the results is not straightforward. 
Another issue with the RPA analysis of multiorbital models is that the investigation of fluctuation-driven phenomena like nematicity requires the inclusion of fluctuations beyond RPA \cite{LauraAlbertoBelenFe, Spindrivennematic} that leads to the definition of a tensorial spin-nematic order parameter. 
In that respect, it has been shown in Ref.~\cite{LauraAlbertoBelenFe, OSSFModel2} that reducing the number of orbitals involved in the calculations by projecting the interaction at low-energy allows to define a minimal model that describes the spin-nematic phase within a simple multiband language while at the same time retaining the orbital information. The low-energy projection, in fact, results in a strong orbital-selectivity of the magnetic excitation with spin-fluctuation along $x/y$ having $yz/xz$ orbital character. 

In this work we perform RPA calculations of the magnetic excitations of the Orbital Selective Spin-Fluctuations (OSSF) model in the tetragonal phase of IBS. 
By comparing our results to analogous microscopic five-orbital calculations we show that the OSSF model reproduces all the relevant features characterizing the RPA spin susceptibilities obtained within multiorbital models. 
The analysis if the SC vertex mediated by OSSF and of the corresponding gap equations, results in anisotropic $s_\pm$ gap functions that can present accidental nodes in agreement with multiorbital calculations and experiments \cite{ReviewHirschfeld, ARPESreview}, as well as a SC $d_{x^2-y^2}$ state nearly degenerate with the $s_\pm$ as previously discussed in e.g. \cite{Graser2009, GapsymmetryHirschfeld2011}. 
The main advantage of the simplified description provided by the OSSF model is that a precise connection between the features of the gap functions and the orbital make-up of the nested Fermi surfaces can be made. Our analysis demonstrates that the degree of orbital nesting is the parameter that controls the modulation and the hierarchy of the SC gaps. This last observation counters the naive expectation of stronger pairing between matching Fermi surfaces and forces us to revise the band nesting paradigm in the light of the orbital degree of freedom.

%%%%%%%%%%%%%%%%%%%%%%%%%%%%%%%%%%%%%%%
\section{The Orbital Selective Spin Fluctuations Model} 
\label{Sec.Model}

\begin{figure}[tbh]
\includegraphics[width=0.42\linewidth]{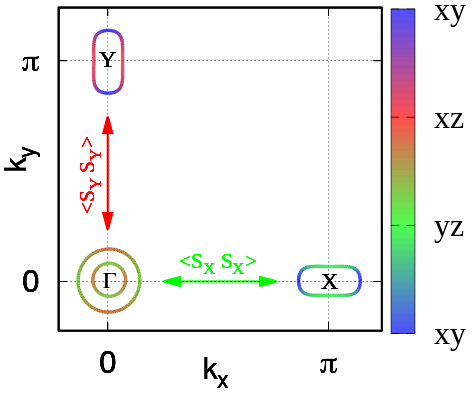}
\caption{Fermi surface for the four-pockets model. The colors represent the main orbital character of the Fermi surface. 
Notice the $yz/xz$ orbital nesting between the inner hole pocket $\Gamma_-$ and the $X/Y$ electron pockets. On the contrary the Fermi surface of the outer hole pocket $\Gamma_+$ presents orbital mismatch with the electron pockets.
The green/red arrows denote the orbital selective spin fluctuations (OSSF), connecting hole and electron pockets at different momenta, see Eqs.~\ref{sx}-\ref{hint_low}.}
\label{fs}
\end{figure}

We start briefly revising the distinctive features of the low-energy OSSF model originally derived in \cite{OSSFModel2}. We consider a general four-pocket model with two hole-pockets at $\G$, $\G_\pm$ and two electron-pockets at $X$ and $Y$, Fig.~\ref{fs}. Mostly three orbitals contribute to the Fermi Surface, $yz$ and $xz$ for the $\Gamma_\pm$ pockets and $xy$, $yz/xz$ for the $X/Y$ pockets. The particular orbital arrangement follows from the space group of the iron-plane \cite{Vafekmodelo}.
A crucial consequence of this orbital composition is that the hole pockets $\Gamma_\pm$ present opposite orbital nesting with the electron pockets, i.e. there is orbital mismatch between the $\Gamma_+$ and $X/Y$ and orbital match between $\Gamma_-$ and $X/Y$. 

The kinetic Hamiltonian is derived adapting the low-energy model considered in~\cite{Vafekmodelo} that accounts for the orbital symmetry of the system. Each pocket is described using a spinor representation in the pseudo-orbital space 
\be
\lb{h0}
H_0^l=\sum_{\bk,\s} \psi^{l,\dagger}_{\bk\s} \hat H_0^l \psi^l_{\bk\s},
\ee
where the spinors are defined as $\psi^{\Gamma}=(c_{yz},c_{xz})$ and $\psi^{X/Y}=(c_{yz/xz},c_{xy})$, $\hat H^l_0= h_0^l\t_0+\vec{h}^l\cdot\vec{\t}$, $l=\G,X,Y$ and the $\t$ matrices represent the pseudo-orbital spin. Diagonalizing $\hat H_0$ we find the dispersion relations $E^{l\pm}=h_0^l\pm h^l$ with $h^l=|\vec{h}^l|$. We introduce the rotation from the orbital to the band basis, 
\be
\lb{hole_fer}
\begin{pmatrix}
 h_+ \\
 h_- \\
\end{pmatrix} = 
\begin{pmatrix}
u_{\Gamma} &-v_{\Gamma}\\
v_{\Gamma} & u_{\Gamma}\\
\end{pmatrix}
\begin{pmatrix}
 c_{yz} \\
 c_{xz} \\
\end{pmatrix} 
\ee
with an analogous expression for the $X/Y$ pockets, provided that the corresponding orbital spinor is used. At  $X/Y$ only the $E^{X/Y+}$ band crosses the Fermi level, so in the following we will use $e_{X/Y}$ for the corresponding fermionic operators dropping the $+$ superscript. The explicit expressions of $(h_0^l, \vec{h}^l)$ that reproduce a four-pocket model as the one shown Fig.~\ref{fs} are detailed in App.~\ref{App.A}. Notice that in order to lift the degeneracy of the inner and outer $xz/yz$ pockets at $\Gamma$ we need to account for the spin-orbit coupling in the Hamiltonian. We added it explicitly by replacing $h^\Gamma\rightarrow \sqrt{(h^\Gamma)^2 + \lambda^2/4}$ in the expression for $E^{\Gamma^\pm}$. The kinetic model we considered here can be easily extended to account for a more general five-pocket model that includes the $xy$ hole-pocket at $(\pi, \pi)$. This band is close to the Fermi level in IBS and it crosses the Fermi level in some specific cases only, e.g. heavily hole doped systems. We will discuss the extension of the OSSF model for a five-pockets model in a future work, however it is worth mentioning that the conclusion discussed here based on our RPA analysis are not expected to change qualitatively once the additional pocket is taken into account.

The interacting Hamiltonian in the spin channel is
\be
\lb{hint}
H_{int}=-1/2\sum_{\bq}U_{\eta\eta'} \vec{S}^\eta (\bq) \cdot \vec{S}^{\eta'} (-\bq).
\ee
$\eta,\eta'=yz,xz,xy$ are orbital indices and $U_{\eta\eta'}\sim U \d_{\eta\eta'} + J_H (1-\delta_{\eta\eta'})$, with $U$ and $J_H$ being the usual Hubbard and Hund couplings. We consider only spin operators with intraorbital character $\vec{S}^\eta (\bq)=\sum_{\bk ss'} \, c^{\eta \dagger}_{\bk s} \vec \s _{s s'}c^\eta_{\bk+\bq s'}$ with $\s_{ss'}$ are the Pauli matrices for the spin operator  being $s,s'$ spin indices. This choice is motivated by the general finding that intraorbital
magnetism is the dominant channel in IBS \cite{Pnictogenheight, Graser2009, Hirschfeldmultiorbital2011, MagneticReview, Nodalintra}. 
Notice that, in order to simplify the notation, we are implicitly assuming any momentum summation normalized via a $1/N$ factor where $N$ is the number of ${\bf k}$-points. 
At low energy we can project out the general interaction, Eq.~\ref{hint}, onto the fermionic excitations defined by the model Eq.~\ref{h0}. By using the rotation to the band basis, Eq.~\ref{hole_fer}, one can then establish a precise correspondence between the orbital and momentum character of the spin operators $\vec S_{X/Y}^\eta\equiv \vec S^\eta ({\bq=\bQ_{X/Y}})$: %
\bea
\lb{sx}
\vec{S}_{X}^{yz} &=& \sum_\bk (u_{\G}h_{+}^{\dagger} 
+ v_{\G} h_{-}^{\dagger} )\, \vec{\s} \, u_{X} e_X \\
\lb{sy}
\vec{S}_{Y}^{xz} &=& \sum_\bk (-v_{\G}h_{+}^{\dagger} 
+ u_{\G} h_{-}^{\dagger} )\, \vec{\s} \, u_{Y} e_Y 
\eea
where we only focus on the spin exchange between hole and electron pockets occurring at momenta $\bq$ near ${\bf Q}_X$ or ${\bf Q}_Y$ and we drop for simplicity the momentum and spin indices of the fermionic operators. The interacting Hamiltonian Eq.~\ref{hint} reduces to 
\be
\lb{hint_low}
H_{int}= - \frac{\tilde{U}}{2} \vec{S}^{yz/xz}_{X/Y} \cdot \vec{S}^{yz/xz}_{X/Y}
\ee
where $\tilde{U}$ is the effective interaction for the low energy model. In the RPA analysis that we perform in the following we will use $\tilde{U}$ close to the critical $U$ value at which the spin RPA susceptibility diverges, leading to the $(\pi,0)$ magnetic order. Notice that Eq.~\ref{hint_low} is the projection of the generic interaction Hamiltonian Eq.~\ref{hint} onto the low-energy model Eq.~\ref{h0}. Such a projection is the key to generate OSSF. In fact, since at low energy the $yz/xz$-fermionic states exist only around $\mathbf{Q}_X/\mathbf{Q}_Y$, it turns out that the spin operators $\vec{S}_{X}^{\eta}$ with $\eta\neq yz$ and $\vec{S}_{Y}^{\eta}$ with $\eta\neq xz$ are absent in Eq.~\ref{hint_low}, and there are no terms involving the Hund's coupling.

The orbital selective character of the low-energy spin-excitations makes the interacting Hamiltonian for the spin-channel, Eq.~\ref{hint_low}, considerably simpler than the one obtained within a five-orbital model (see e.g. \cite{Graser2009}). As a matter of fact,  Eq.~\ref{hint_low}, while retaining the orbital dependence of the spin excitations, does not acquire a complex tensorial form and is instead formally equivalent to the spin-spin interacting Hamiltonian written in the band-basis $H^{band}_{int} \sim \ - \tilde{U} \ \vec{S}_{X/Y} \cdot \vec{S}_{X/Y}$ (see e.g. \cite{PreemptiveFernandes2012}).
This implies that one can analyze the spin-nematic phase following the same strategy of \cite{PreemptiveFernandes2012}, in which the nematic instability has been studied within an effective action derivation as a precursor effect of magnetism. 
The consequences of the inclusion of the orbital degree of freedom within a spin-nematic action have been widely discussed especially in the analysis of the nematic phase of FeSe \cite{ OSSFModel2, OSSFModel1, OSSFNematic, Paper1}. 

In what follows we focus on the analysis of the superconductivity mediated by OSSF in the tetragonal phase for a generic four-pocket model representative of IBS. We will show that the orbital selectivity of the spin-fluctuations makes the RPA treatment extremely simple with respect to multiorbital calculations. Given the intraorbital scalar character of the OSSF, the analysis turns out to be mathematically equivalent to the study of single-band systems, while retaining all the multiband and multiorbital information.

%%%%%%%%%%%%%%%%%%%%%%%%%%%%%%%%%%%%%%%
\section{Results}
\label{Sec:RPA}

Within standard RPA analysis the pairing is assumed to be mediated by spin and charge fluctuations \cite{Kuroki2008, Graser2009, GapsymmetryHirschfeld2011, Acommonthread}. It has been shown that for IBS the charge susceptibility is more than one order of magnitude smaller than the spin susceptibility (see e.g. \cite{Graser2009}), therefore hereafter we focus on the spin channel only. 

The spin susceptibility for a generic multiorbital system is a four orbital indices tensor, $\chi^{\eta \eta'}_{\delta \delta'} (\mathbf{q},\Omega)$. This is obtained from the analytical continuation  $i\Omega_{m} \rightarrow \Omega + i0^{+}$ of the Matsubara spin-spin correlation function  
\begin{equation}
\chi^{\eta \eta'}_{\delta \delta'}(\mathbf{q}, i\Omega_{m}) = \int_{0}^{\beta} d\tau \, e^{i \Omega_{m} \tau} \,  \bigg\langle T_{\tau}  \vec{S}^{\eta\delta} (\mathbf{q}, \tau) \vec{S}^{\eta' \delta'} (\mathbf{-q}, 0)  \bigg\rangle 
\label{SFMulti1}
\end{equation}
where $\mathbf{q}$ is the momentum vector, $\beta = 1/k_BT$ is the inverse temperature, $\tau$ is the imaginary time and $\Omega_m = 2m\pi k_BT$ is the bosonic Matsubara frequency. The spin operator in the orbital space for the $\eta,\delta$ orbitals is defined as
$\vec{S}^{\eta\delta} (\mathbf{q}, \tau)=\sum_{\mathbf{k} ss'} \, c^{\eta \dagger}_{\mathbf{k} s} \vec \sigma _{s s'}c^\delta_{\mathbf{k}+\mathbf{q} s'}$
with $ \vec \sigma _{s s'}$ the Pauli matrices for the spin operator. Using this explicit definition of $\vec{S}^{\eta\delta}$ and applying the Wick's theorem to Eq.~\ref{SFMulti1} the non-interacting spin susceptibility can be rewritten as
\begin{equation} 
\chi^{\eta \eta'}_{\delta \delta'}(\mathbf{q}, \Omega_m) = - \frac{1}{\beta} \sum_{\mathbf{k},n} Tr \left[G^{\delta \eta}  (\mathbf{k},i\omega_n)
 G^{\delta' \eta'} (\mathbf{k}+\mathbf{q},i\omega_n+i\Omega_m)  \right] 
\label{SFMulti3}
\end{equation}
where the spectral representation of the Green's function is given by the rotation to the orbital basis of the non-interacting Green's function in the band basis 
\begin{equation}
G^{\delta \eta}  (\mathbf{k},i\omega_n) = \sum_m \frac{a_m^\delta(\mathbf{k})a_m^{\eta*}(\mathbf{k})}{i\omega_n -E_m(\mathbf{k})}
\label{SFMulti4}
\end{equation}
where $\omega_n = (2n+1)\pi k_BT$ is the fermionic Matsubara frequency and $a_m^\eta(\mathbf{k})$ the matrix elements connecting the orbital ($\eta$) and the band space ($m$) determined by diagonalization of the tight-binding Hamiltonian.
Performing the Matsubara frequency summation and setting $\Omega \rightarrow 0$, the static spin susceptibility reads 
\begin{equation}
\chi^{\eta \eta'}_{\delta \delta'}(\mathbf{q}) =  - \sum_{\mathbf{k},m n} \frac{a_m^\delta(\mathbf{k})a_m^{\eta*}(\mathbf{k})a_n^{\eta'*}(\mathbf{k}+\mathbf{q})a_n^{\delta'}(\mathbf{k}+\mathbf{q})}{E_{n}(\mathbf{k}+\mathbf{q})-E_{m}(\mathbf{k})} {f(E_{n}(\mathbf{k}+\mathbf{q}))-f(E_m(\mathbf{k}))}
\label{SFMulti5}
\end{equation}
with $f(E_m(\mathbf{k}))$ the Fermi distribution function. The RPA spin-fluctuation is given in the form of a Dyson-type equation with the spin interaction $\hat{U}_S$ defined in terms of the multiorbital interaction parameters $U,U',J_H$ \cite{Kuroki2008, Graser2009}. Analogously, the singlet pairing vertex driven by spin-fluctuations can be computed on the low-energy sector in terms of the RPA spin susceptibility \cite{Graser2009, Roleofvertex}. 
The variety of diagrams contributing to the SC vertex is large given the number of orbitals included making unfeasible to draw the possible Feynman diagrams up to orders larger than one.
The gap equation for the multiorbital model can be computed numerically by taking into account the singlet pairing vertex as an eigenvalue problem in which the largest eigenvalue leads to the highest transition temperature and its eigenfunction determines the symmetry of the gap (see e.g. \cite{Graser2009, GapsymmetryHirschfeld2011, Acommonthread, Kemper2010}). An anisotropic sign changing s-wave, $s_\pm$, is found as the dominant symmetry for system parameters compatible with moderately doped IBS, in agreement with experiments, e.g. \cite{NMRgap, NMRgap2, ResonancepeakNSIBS,ARPESreview}. A nearly degenerate  $d_{x^2-y^2}$ state has been discussed in \cite{Graser2009, GapsymmetryHirschfeld2011} and could be relevant to explain Raman experiments in K-doped BaFe$_2$As$_2$ \cite{Loidl_PRL2013, HaiHu_PRX2014,Blumberg_PRB2017, Hackl_QM2018}, CaKFe$_4$As$_4$ \cite{Hack_PRB2018} and 
(Li$_{1-x}$Fe$_x$)OHFeSe \cite{Hackl_PRL2020}.
\subsection{Magnetic excitations in the OSSF model: RPA analysis} \label{Sec.SFRPA}

Within the OSSF model, the situation is substantially simplified as compared with the five-orbital RPA approach due to the orbital-selectivity of the spin-fluctuations.
By assuming the spin-operator to be intraorbital the spin-susceptibility of Eq.~\ref{SFMulti1} reduces to a two-orbital indices matrix
\begin{equation}
\chi^{\eta \eta'}(\mathbf{q}, i\Omega_{m}) = \int_{0}^{\beta} d\tau \, e^{i \Omega_{m} \tau} \,  \bigg\langle T_{\tau}  \vec{S}^{\eta} (\mathbf{q}, \tau) \vec{S}^{\eta'} (\mathbf{-q}, 0)  \bigg\rangle 
\label{SFOSSF1}
\end{equation}
The low-energy projection further simplified the spin-susceptibility structure as the low-energy states are defined only around high symmetry points and have a well defined orbital character described by Eq.~\ref{h0}. 
As a consequence, also the Green's functions are defined only for  $l= \Gamma, X, Y$ as $G^{l}(\mathbf{k},i\omega_n)={\hat{\cal U}^l}(\mathbf{k},i\omega_n)diag (g^{l_+}(\mathbf{k},i\omega_{n}),g^{l_-}(\mathbf{k},i\omega_{n})){\hat{\cal U}^{l^{-1}}}(\mathbf{k},i\omega_n)$. Here $\hat{\cal U}^l$ are the matrices that diagonalize the $l$-Hamiltonian and $g^{l\pm} (\mathbf{k}, i\omega_n)= (i\omega_n-E^{l\pm}_\mathbf{k})^{-1}$ the Green's functions in the band basis. Substituting the intraorbital spin operator $\vec{S}^{\eta} (\mathbf{q}, \tau)$ and applying Wick’s theorem to Eq.~\ref{SFOSSF1}, the intraorbital spin susceptibility in the low-energy projection can be read as
\begin{equation} 
\chi^{l l'}(\mathbf{q}, i\Omega_m) = - \frac{1}{\beta} \sum_{\mathbf{k},n} Tr \left[\hat{G}^l(\mathbf{k},i\omega_n)
 \hat G^{l'} (\mathbf{k}+\mathbf{q},i\omega_n+i\Omega_m)  \right] \,.
 \label{SFOSSF2}
\end{equation}
Eq.~\ref{SFOSSF2} represents the spin susceptibility between two pockets $l$ and $l^{'}$ and depends on the transferred momentum $\mathbf{q}=\mathbf{k'}-\mathbf{k}$ and the external frequency $\Omega$. 
Performing the Matsubara frequency summation and setting $\Omega \rightarrow 0$ in Eq.~\ref{SFOSSF2} \cite{app}, 
we find the static susceptibility for the $l_{\pm}$ $l^{'}_{\pm}$ pockets in terms of the Fermi distribution function $f(E^{l_\pm}_{\mathbf{k}})$ and  $|(u/v)^{l}(\mathbf{k})|$ that are the elements of the rotational matrix $\hat{\cal U}^l$ connecting the orbital and the band space 
\begin{equation}
\chi^{l_\pm l'_\pm}_{\eta}(\mathbf{q}) = \sum_{\mathbf{k}} \frac{f(E^{l'_\pm}_{\mathbf{k}+\mathbf{q}})-f(E^{l_\pm}_{\mathbf{k}})}{E^{l'_\pm}_{\mathbf{k}+\mathbf{q}}-E^{l_\pm}_{\mathbf{k}}} |(u/v)^l_{\mathbf{k}}|^2 |(u/v)^{l'}_{\mathbf{k}+\mathbf{q}}|^2 \,.
\label{SFOSSF3}
\end{equation}
The resulting static susceptibility Eq.~\ref{SFOSSF3}, although formally similar to the multiorbital spin-susceptibility in Eq.~\ref{SFMulti5}, is much simpler due to the orbital selectivity of the spin fluctuations. In fact, within the OSSF the two most relevant spin susceptibilities for a four-pocket model only involve the $yz$ orbital coming from the interaction between the $\Gamma_\pm$ holes with the $X$ electron pockets, and the $xz$ orbital coming from the $\Gamma_\pm$ holes with the $Y$ electron pockets near $\mathbf{Q}_X$ and $\mathbf{Q}_Y$ respectively. To better compare with the results obtained within five-orbital model calculations, in what follows we account also for the the spin susceptibility centered around $\mathbf Q_M=(\pi,\pi)$ that describes the spin exchange between the $X$ and $Y$ electron pockets and, at low-energy, involves the $xy$ orbital only.

The RPA spin susceptibilities are obtained in the form of Dyson-type equations. The results of the resummation reads  
\begin{equation}
\chi^{l_\pm l'_\pm}_{\eta \rm{RPA}}(\mathbf{q})  = \frac{\chi_{\eta}^{l_\pm l'_\pm}(\mathbf{q})}{1- \tilde{U}\chi_{\eta}^{l_\pm l'_\pm}(\mathbf{q})}
\label{SFOSSF4}
\end{equation}
with $\chi_{\eta}^{l_\pm l'_\pm}(\mathbf{q})$ the non-interacting spin susceptibility given by Eq.~\ref{SFOSSF3}. The set of orbital-selective RPA suscpetibility for the different pockets are given in App.~\ref{App.B}. Notice that, within our model Eq.~\ref{hint_low}, both the low-energy effective coupling $\tilde{U}$  and the OSSF $\chi_{\eta}^{l_\pm l'_\pm}$ are intraorbital and have scalar character. As a consequence, the RPA spin susceptibility given by Eq.~\ref{SFOSSF4} is straightforward and inherits the orbital selectivity and scalar form of the OSSF. As we will discuss in the following Sections~\ref{Sec.VertexRPAandGAP} and ~\ref{SubSec.GAP} this aspect allows us to derive analytical expressions for the pairing vertex and SC gaps.

\begin{figure}[tbh]
\includegraphics[width=0.97\linewidth]{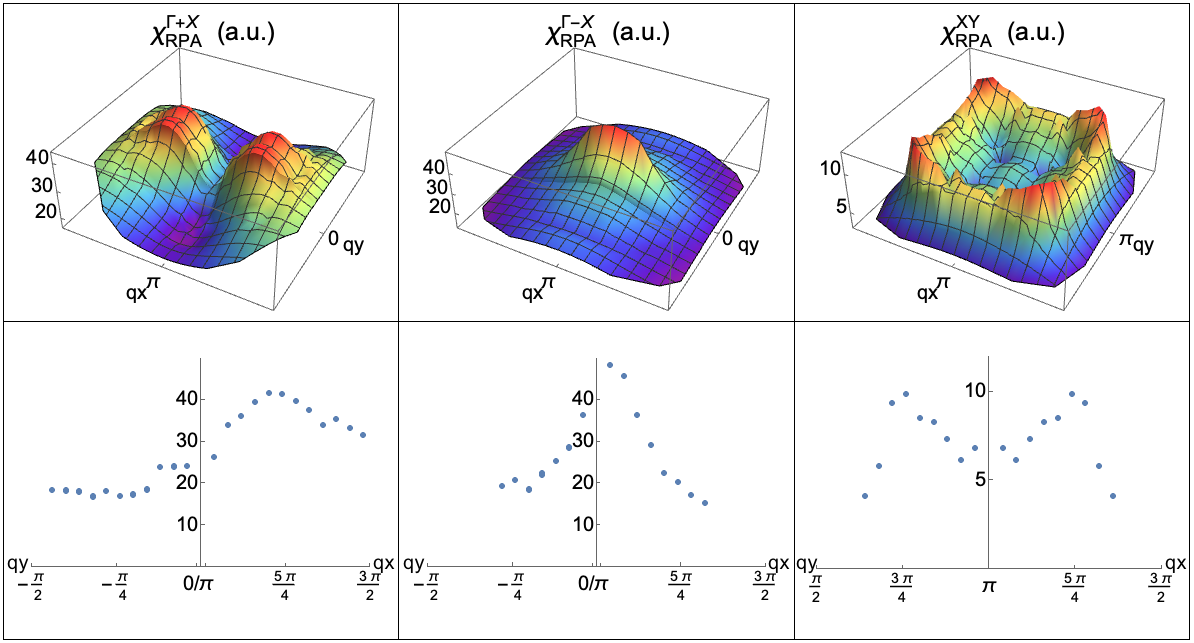}
\caption{RPA spin susceptibility for a four-pocket model in the tetragonal phase. 3D maps (upper panels) and 2D cuts (bottom panels) of the RPA susceptibilities shown in the $q_x-q_y$ space around the high symmetry points $X$ for the hole-electron sector and $M$ for the electron-electron sector. The orbital composition of the nested bands is at the origin of the momentum dependence of the RPA spin susceptibilities that present commensurate/incommensurate peaks depending on the orbital match/mismatch of the nested Fermi surfaces. The band parameters used in the calculations are detailed in App.~\ref{App.D}, the interaction is fixed at $\tilde{U}$ = 1~eV and the temperature is $0.02$eV.}
\label{fig:122RPA}
\end{figure}

%\textit{Numerical estimate}
To get insight into the previous result, we perform a numerical estimate for the RPA spin susceptibility given by Eq.~\ref{SFOSSF4} for the four-pocket model in Fig.~\ref{fs}. 
We show  $\chi^{\Gamma_+ X}_{yz \rm{RPA}}(\mathbf{q})$, $\chi^{\Gamma_- X}_{yz \rm{RPA}}(\mathbf{q})$ and $\chi^{XY}_{xy \rm{RPA}}(\mathbf{q})$ in Fig.~\ref{fig:122RPA}. 
In the upper panel we show 3D color maps in $(q_x,q_y)$. 
The bottom panels show the 2D cuts along $q_x/q_y$ centered in $\mathbf{Q}_X=(\pi,0)$ for the electron-hole spin susceptibilities and $\mathbf{Q}_M=(\pi,\pi)$ for the electron-electron one. This representation allows to easily compare the relative weight of the different susceptibilities.
Notice that the contributions of the $Y$ pocket (not shown) are equivalent to those for the $X$ pockets with a $\pi/2$ rotation, since in the tetragonal phase the susceptibility is isotropic in both directions. 

From Fig.~\ref{fig:122RPA} we can highlight two main results:

(i) The orbital-selective RPA spin susceptibilities around $\mathbf{Q}_X$ and $\mathbf{Q}_M$ show a clear momentum-dependent structure of the peaks. This can be explained due to the degree of orbital nesting between pockets. The orbital nesting indicates the relative orbital composition between the two pockets involved in the spin-exchange mechanism. 
In Fig.~\ref{fig:122RPA} we can see that when there is an orbital mismatch, as is the case of $\Gamma_+$ and $X$ pockets, the spin susceptibility develops two incommensurate peaks around $\mathbf{Q}_X$. In contrast, if there is an orbital match between pockets, i.e. the case of $\Gamma_-$ and $X$, the spin susceptibility develops a single commensurate peak at the $\mathbf{Q}_X=(\pi,0)$. For the $\chi^{XY}_{xy \rm{RPA}}(\mathbf{q})$ susceptibility there is a total mismatch between the $xy$ orbital of the electrons pockets. Thus, the spin susceptibility is totally incommensurate and develops four symmetric peaks that correspond to the overlaps of the $xy$ orbital contribution around the $(\pi,\pi)$ point. 

(ii) The main contribution to the spin susceptibility comes from the $\mathbf{Q}_{X/Y}$ spin-mode, between the $\Gamma_-$ and the $X/Y$ pockets. This is a consequence of the orbital composition of the nested bands. The Fermi surface of $\Gamma_-$ and and $X/Y$ are, in fact, characterized by fully matching orbitals.

%Comparacion con multiorbital de MJ
We compare our results with numerical calculations of the RPA spin susceptibility for a five-orbital model adapted from \cite{Graser2009}.
By tuning the filling and the crystal field we can consider separately two different cases: the first one corresponds to a four-pocket model with better band nesting between the $\Gamma_-$ and the electron pockets, and the second case to a four-pocket model with better band nesting between the $\Gamma_+$ and the $X/Y$ pockets. We obtain for the first case a commensurability of the RPA spin susceptibility at the $\mathbf{Q}_X=(\pi,0)$, whereas in the second case we get non-commensurate peaks around $\mathbf Q_X=(\pi,0)$. Therefore, the same orbital modulation for the momentum dependence of the RPA spin susceptibility is obtained within the OSSF and the multiorbital models. 
We also compute the RPA spin susceptibility coming from the electron-electron sector within the five-orbital model. As expected, we find that the contribution from this sector is negligible in comparison with the one for the hole-electron sector \cite{ReviewHirschfeld}. 
It is worth noticing that, while the existence of a correlation between the orbital make-up of the Fermi surface and the momentum-dependent structure of the RPA spin excitation has already been highlighted within multiorbital models (e.g. \cite{GapanisotropycomparitionRPA}), the explicit link between orbital nesting and momentum dependence of the spin-susceptibility is made transparent within the RPA analysis of the OSSF model.
\begin{figure}[tbh]
\centering     
\subfloat[OSSF model]{\includegraphics[width=75mm]{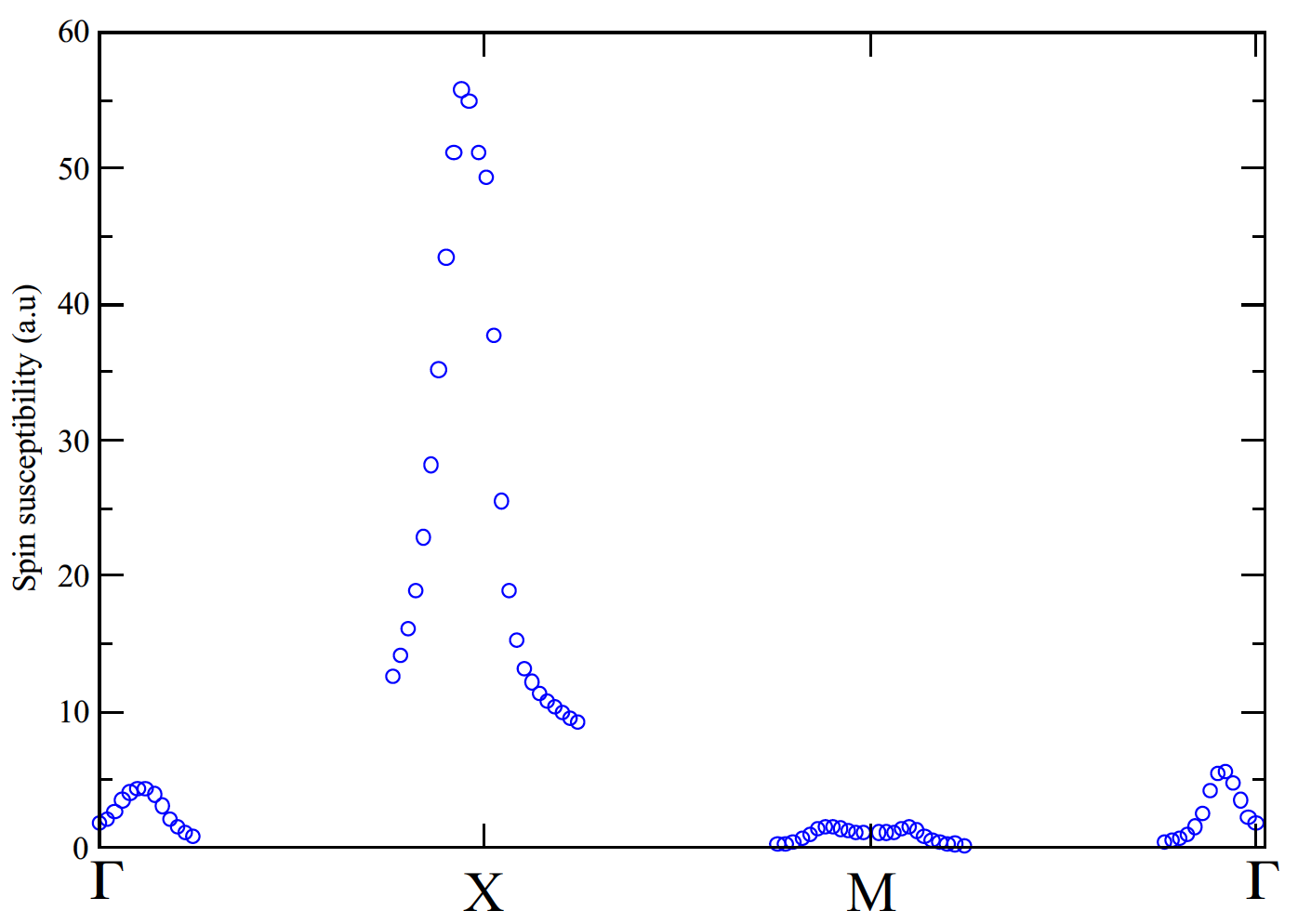}} 
\subfloat[Five-orbital model]{\includegraphics[width=75mm]{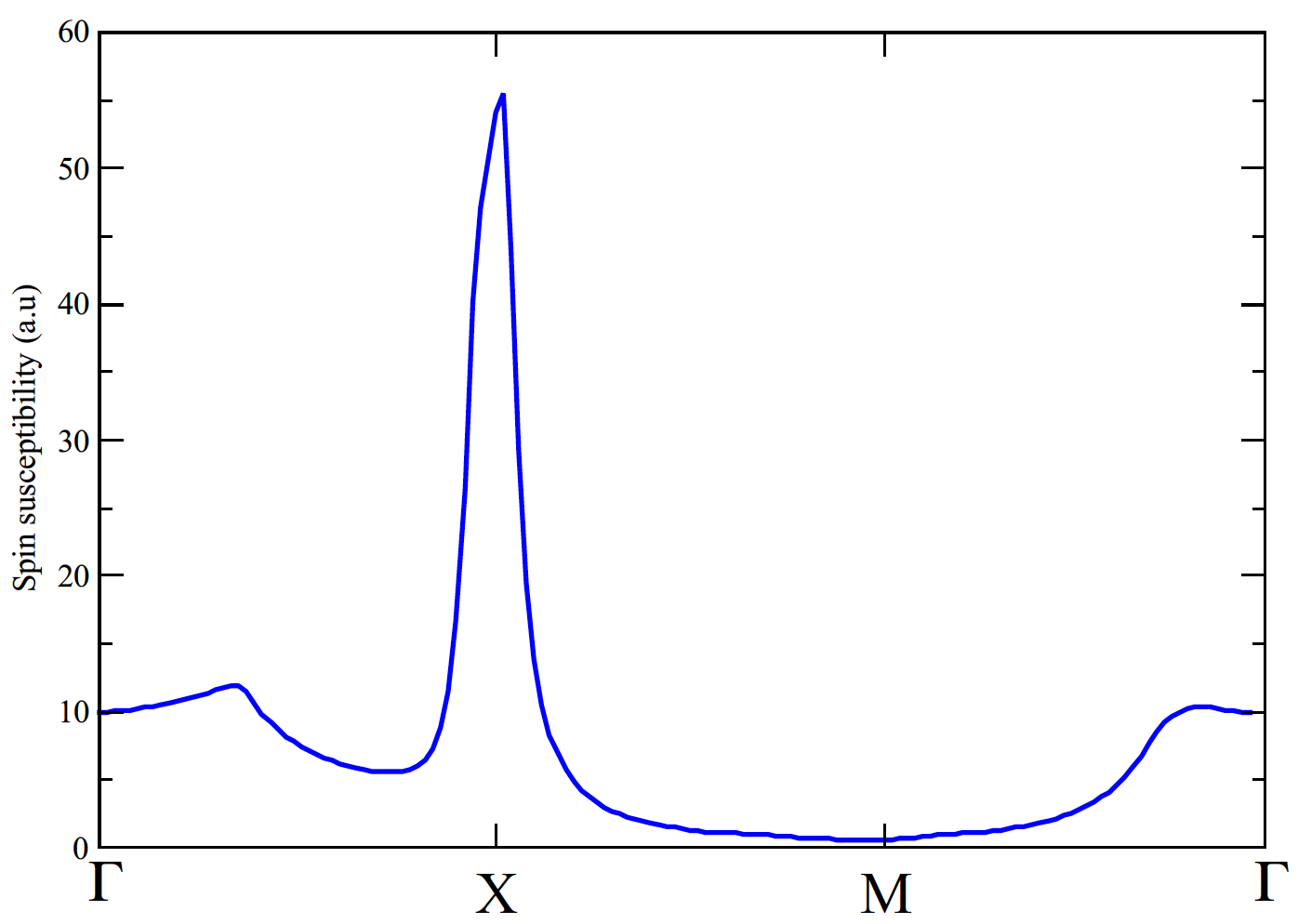}}
\caption{Total RPA spin susceptibility for the four-pocket model. We show the cuts along the high symmetry directions $\Gamma X M \Gamma$ of the RPA spin susceptibility obtaines within (a) the OSSF model and the (b) the five-orbital model. Temperature is fixed to $T=0.02$eV in both panels. We set the effective low-energy interaction as $\tilde{U}$ = 1~eV in (a), while the intraorbital and the interorbital onsite interactions for the five-orbital models are set to $U$ = 1.2~eV, $J_H = 0.25U$ and $U'= U-2J_H$ in (b). The RPA susceptibility for the five-orbital model is computed for any momentum ${\bf q}$ as it follows from a full-band calculation. On the other hand, the OSSF model is a low-energy  model, thus provides information of the bands at the Fermi level (Fermi pockets) only. This implies that not all possible values of ${\bf q}$ are allowed. The RPA susceptibility is this case is well defined around the high symmetry points $\Gamma, X, M$ only as one can see in panel (a). Nonetheless we find a remarkable qualitative agreement within the low-energy calculation and the five-orbital one.}
\label{fig:CaminoGXMG}
\end{figure}

In Fig.~\ref{fig:CaminoGXMG} we compare the cuts along the main symmetry directions of the RPA spin susceptibility (including the intraband ones) computed within the OSSF model and the multiorbital model. The calculation performed within the OSSF model reproduces remarkably well the overall momentum-dependence of the spin-spectrum as well as the relative height and width of the various peaks.
This comparative analysis proves that we can obtain a reliable description of the spin spectrum within the OSSF model of Eq.~\ref{SFOSSF3}, without dealing with the tensorial form of the spin susceptibility given in the five-orbital model, Eq.~\ref{SFMulti5}. 

\subsection{Superconductivity mediated by OSSF} \label{Sec.VertexRPAandGAP}

\begin{figure}[tbh]
\centering
{\includegraphics[width=0.99\linewidth]{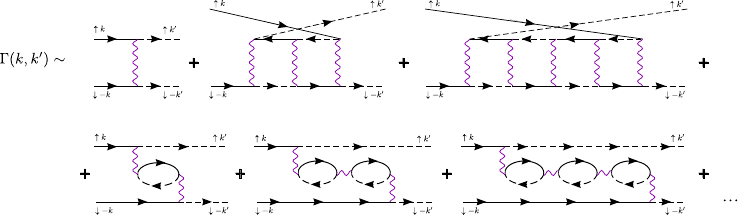}}
\caption{Pairing vertex in the random phase approximation up to fourth order within the OSSF model.}
\label{fig:Vertex}
\end{figure}

In Fig.~\ref{fig:Vertex} we show the RPA leading diagrams for the SC vertex of electrons of opposite spin and momentum. Within the OSSF model we can explicitly draw the Feynman's diagrams up to the a finite order in $\tilde{U}$. This is possible given the intraorbital scalar character of the low-energy orbital-selective spin-susceptibility that allows us to easily perform the diagrammatic expansion explicitly. Notice that the expansion is instead practically unfeasible within the five-orbital model due to the complex tensorial structure of the pairing vertex.  

As a consequence of the projection to a constrained orbital space within the OSSF model, the diagrammatic expansion for the RPA pairing vertex is formally equivalent to the one for the band-basis model which, however, does not contain the orbital information of the spin-fluctuations exchange. Within the multiband model, the RPA pairing vertex is composed by the exchange of spin fluctuations connecting an electron pocket with a hole pocket. On the other hand, within the OSSF model the essential difference is that the low energy exchanged spin fluctuations connect a hole with an electron pocket with the {\it same orbital content} $yz$ or $xz$. In this way, we retain the simplicity of the analysis of the Feynman's diagrams within the multiband model and, at the same time, we account for the orbital degree of freedom of the system. 

The leading RPA diagrams for the vertex, Fig.~\ref {fig:Vertex}, can be written as
\begin{equation}
\Gamma^{l_\pm l'_\pm}_{\eta}(\mathbf{q}) =  
\tilde{U} 
+   \frac{ \tilde{U}^2  \chi_{\eta}^{l_\pm l'_\pm}(\mathbf{q})}{1-\tilde{U} \chi_{\eta}^{l_\pm l'_\pm}(\mathbf{q})}
+ \frac{\tilde{U}^3  \chi_{\eta}^{{l_\pm l'_\pm}^2}(\mathbf{q})}{1-\tilde{U}^{2} \chi_{\eta}^{{l_\pm l'_\pm}^2}(\mathbf{q})} 
\label{SFOSSF5}
\end{equation}
where $\mathbf{q}=\mathbf{k'}-\mathbf{k}$ is the transferred momentum, $\tilde{U}$ is the intraorbital effective coupling and $\chi_{\eta}^{l_\pm l'_\pm}(\mathbf{q})$ the intraorbital susceptibility given in formula Eq.~\ref{SFOSSF3}. 
As we are considering the spin channel only, the SC vertex is proportional to the RPA spin susceptibility and preserves the orbital and momentum dependencies and the physical properties discussed in the previous section. For instance, we get the same criterion of commensurability or incommensurability depending on the orbital matching between pockets. In the same way, we also obtain that the dominant contribution to the RPA pairing vertex is given by the spin-fluctuations exchange between hole and electron pockets, being the largest between pockets having matching orbital, i.e. $\Gamma_- - X/Y$.

%%%%%%%%%%%%%%%%%%%%%%%%%%
\subsection{Superconducting gaps} \label{SubSec.GAP}
\begin{figure}[tbh]
\centering  
\subfloat[]{\includegraphics[scale=0.4]{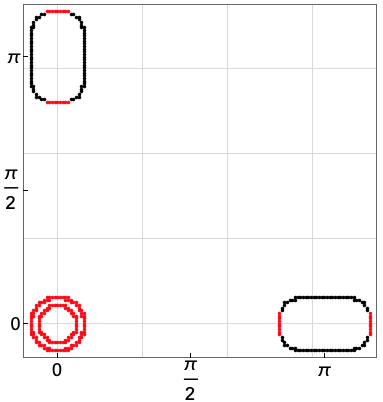}} 
\subfloat[]{\includegraphics[scale=0.4]{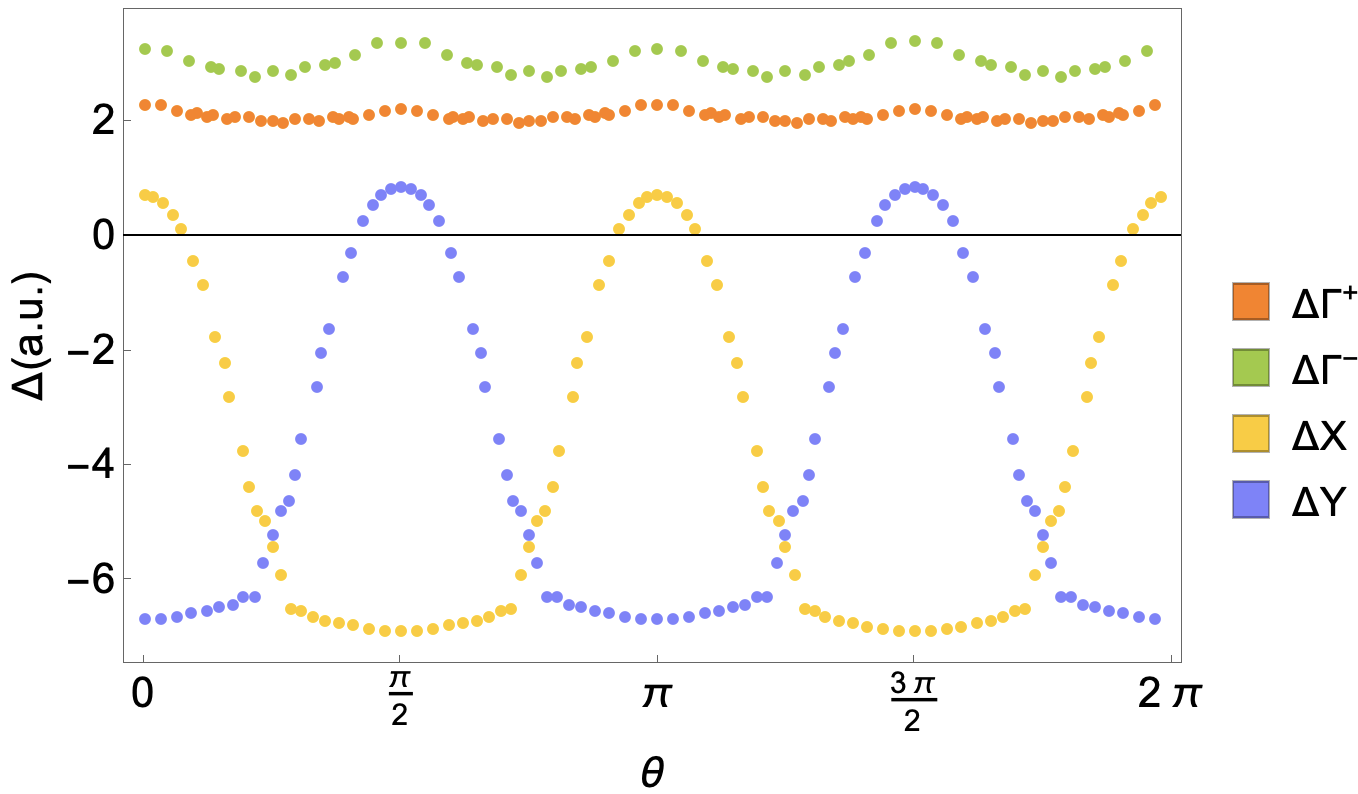}}
\caption{RPA gap functions within the OSSF model (a) plotted on the Fermi surface pockets (red circles for positive and black circles for negative gaps), (b) plotted as a function of angle from $0$ to $2\pi$. The SC gap functions present $s_\pm$ symmetry with angular modulations controlled by the orbital composition of the nested bands. The hierarchy of the band gaps follows from the orbital nesting properties of the matching Fermi surface pockets. Band parameters are detailed in App.~\ref{App.D}, $\tilde{U} = 1.5$ eV.}
\label{fig:gapnodo}
\end{figure}

We solve the BCS gap equations mediated by the OSSF for a four pocket model for IBS in the tetragonal phase. In order to better highlight the effect of the orbital nesting between the hole and electron nested pockets we consider a case in which the two hole pockets are almost equivalent in size \cite{app}, see Fig. \ref{fig:gapnodo}. In this way the degree of band nesting between the hole and electron pockets is the same for the inner and outer hole pockets, but the orbital matching condition determined by the space group of the iron-plane \cite{Vafekmodelo} is different for $\Gamma_\pm$.
For simplicity we show here the equations for the $yz$ and $xz$ components only. However, we have numerically solved the full orbital system in which we also consider the SC vertex mediated by the exchange of spin-fluctuations between the $X-Y$ pockets, see App.~\ref{App.C}. 
The pairing Hamiltonian for the $yz$ and $xz$ orbital contributions reads
\begin{eqnarray}
H^{pairing}_{yz,xz} &=& - \sum_{\mathbf{k},\mathbf{k'}} \Gamma^{\Gamma_+ X}_{yz \ \mathbf{k}\mathbf{k'}} \left[(u^\Gamma_{\mathbf{k}})^2 h^{+\dagger}_{\mathbf{k}} h^{+\dagger}_{\mathbf{-k}} (u^X_{\mathbf{k}})^2 e^{X}_{\mathbf{k'}} e^{X}_{\mathbf{-k'}} \right] - \sum_{\mathbf{k},\mathbf{k'}} \Gamma^{\Gamma_- X}_{yz \ \mathbf{k}\mathbf{k'}} \left[(v^{\Gamma}_{\mathbf{k}})^2 h^{-\dagger}_{\mathbf{k}} h^{-\dagger}_{\mathbf{-k}} (u^X_{\mathbf{k}})^2 e^{X}_{\mathbf{k'}} e^{X}_{\mathbf{-k'}} \right] \nn \\
&-& \sum_{\mathbf{k},\mathbf{k'}} \Gamma^{\Gamma_+ Y}_{xz \ \mathbf{k}\mathbf{k'}} \left[(v^\Gamma_{\mathbf{k}})^2 h^{+\dagger}_{\mathbf{k}} h^{+\dagger}_{\mathbf{-k}} (u^Y_{\mathbf{k}})^2 e^{Y}_{\mathbf{k'}} e^{Y}_{\mathbf{-k'}} \right] - \sum_{\mathbf{k},\mathbf{k'}} \Gamma^{\Gamma_- Y}_{xz \ \mathbf{k}\mathbf{k'}} \left[(u^{\Gamma}_{\mathbf{k}})^2 h^{-\dagger}_{\mathbf{k}} h^{-\dagger}_{\mathbf{-k}} (u^Y_{\mathbf{k}})^2 e^{Y}_{\mathbf{k'}} e^{Y}_{\mathbf{-k'}}  \right] + h.c. \nn \\
\label{BCS1}
\end{eqnarray}
where $\Gamma_{\eta \ \mathbf{k}\mathbf{k'}}^{l_\pm l'_\pm}$ with $\eta=yz/xz$ is the RPA pairing vertex given by Eq.~\ref{SFOSSF5} for the different pockets ${l_\pm l'_\pm}$ and $(u^l_\mathbf{k})^2$ and $(v^l_\mathbf{k})^2$ are the coherence factors that connect the orbital and the band basis and account for the pockets orbital character.

The pairing Hamiltonian given by Eq.~\ref{BCS1} is solved in the mean field approximation by defining the orbital dependent SC order parameters for the hole sector 
($\Delta_{yz}^{\Gamma_+ X},\Delta_{yz}^{\Gamma_-X},\Delta_{xz}^{\Gamma_+Y},\Delta_{xz}^{\Gamma_- Y}$) 
and for the electron sector 
($\Delta_{yz}^{X\Gamma_+},\Delta_{yz}^{X\Gamma_-},\Delta_{xz}^{Y \Gamma_+},\Delta_{xz}^{Y\Gamma_-}$).
The resulting linearized gap equations at $T=0$ read as
\begin{eqnarray}
\Delta^{\Gamma_+X}_{yz\mathbf{k'}}&=&-\sum_{k} \frac{\Gamma^{\Gamma_+ X}_{yz\mathbf{k}\mathbf{k'}} }{|v^X_{F\mathbf{k}}|} (u^{X}_{\mathbf{k}})^{4} \left[ 
\Delta^{X\Gamma_+}_{yz\mathbf{k}} + 
\Delta^{X\Gamma_-}_{yz\mathbf{k}} \right]  \label{BCS2.1}  \\
\Delta^{\Gamma_- X}_{yz\mathbf{k'}} &=& - \sum_{k} \frac{\Gamma^{\Gamma_- X}_{yz\mathbf{k}\mathbf{k'}} }{|v^X_{F\mathbf{k}}|} (u^{X}_{\mathbf{k}})^{4} \left[ 
\Delta^{X \Gamma_+}_{yz\mathbf{k}} +
\Delta^{X \Gamma_-}_{yz\mathbf{k}} \right]  \label{BCS2.2}  \\
\Delta^{\Gamma_+ Y}_{xz\mathbf{k'}} &=& - \sum_{k} \frac{\Gamma^{\Gamma_+ Y}_{xz\mathbf{k}\mathbf{k'}} }{|v^Y_{F\mathbf{k}}|} (u^{Y}_{\mathbf{k}})^{4} \left[ 
\Delta^{Y \Gamma_+}_{xz\mathbf{k}} + 
\Delta^{Y \Gamma_-}_{xz\mathbf{k}} \right]  \label{BCS2.3}  \\
\Delta^{\Gamma_- Y}_{xz\mathbf{k'}} &=&  - \sum_{k} \frac{\Gamma^{\Gamma_- Y}_{xz\mathbf{k}\mathbf{k'}} }{|v^Y_{F\mathbf{k}}|} (u^{Y}_{\mathbf{k}})^{4}  \left[
\Delta^{Y \Gamma_+}_{xz\mathbf{k}} + 
\Delta^{Y \Gamma_-}_{xz\mathbf{k}} \right]  \label{BCS2.4} \\
\Delta^{X \Gamma_+}_{yz\mathbf{k'}} &=& - \sum_{k} \frac{\Gamma^{\Gamma_+ X}_{yz\mathbf{k}\mathbf{k'}} }{|v^{h_+}_{F\mathbf{k}}|} (u^{\Gamma}_{\mathbf{k}})^2 \left[ (u^{\Gamma}_{\mathbf{k}})^2 \Delta^{\Gamma_+ X}_{yz\mathbf{k}} + (v^{\Gamma}_{\mathbf{k}})^2 \Delta^{\Gamma_+ Y}_{xz\mathbf{k}} \right]  \label{BCS2.5}  \\
\Delta^{X \Gamma_-}_{yz\mathbf{k'}} &=&  - \sum_{k} \frac{\Gamma^{\Gamma_- X}_{yz\mathbf{k}\mathbf{k'}} }{|v^{h_-}_{F\mathbf{k}}|} (v^{\Gamma}_{\mathbf{k}})^2 \left[ (v^{\Gamma}_{\mathbf{k}})^2 \Delta^{\Gamma_- X}_{yz\mathbf{k}} + (u^{\Gamma}_{\mathbf{k}})^2 \Delta^{\Gamma_-Y}_{xz\mathbf{k}} \right] \label{BCS2.6}  \\
\Delta^{Y \Gamma_+}_{xz\mathbf{k'}} &=&  -\sum_{k} \frac{\Gamma^{\Gamma_+ Y}_{xz\mathbf{k}\mathbf{k'}} }{|v^{h_+}_{F\mathbf{k}}|} (v^{\Gamma}_{\mathbf{k}})^2 \left[ (u^{\Gamma}_{\mathbf{k}})^2 \Delta^{\Gamma_+ X}_{yz\mathbf{k}} + (v^{\Gamma}_{\mathbf{k}})^2 \Delta^{\Gamma_+Y}_{xz\mathbf{k}} \right] \label{BCS2.7}   \\
\Delta^{Y \Gamma_-}_{xz\mathbf{k'}} &=&   - \sum_{k} \frac{\Gamma^{\Gamma_- Y}_{xz\mathbf{k}\mathbf{k'}} }{|v^{h_-}_{F\mathbf{k}}|} (u^{\Gamma}_{\mathbf{k}})^2 \left[ (v^{\Gamma}_{\mathbf{k}})^2 \Delta^{\Gamma_-X}_{yz\mathbf{k}} + (u^{\Gamma}_{\mathbf{k}})^2 \Delta^{\Gamma_-Y}_{xz\mathbf{k}} \right] \label{BCS2.8} \end{eqnarray}
with $v^l_{F\mathbf{k}}=\partial_\mathbf{k}(\epsilon_{\mathbf{k}}^{l\pm })$ the Fermi velocity for the pocket $l$. Eqs.~\ref{BCS2.1}\,-\,\ref{BCS2.8} represent the orbital components for each gap. Then, we define the total low-energy band gaps $\Delta_\mathbf{k}^{l}$ as
\begin{eqnarray}
\Delta_\mathbf{k}^{\Gamma_+} &=& (u^{\Gamma}_{\mathbf{k}})^2 \Delta^{\Gamma_+ X}_{yz\mathbf{k}} + (v^{\Gamma}_{\mathbf{k}})^2 \Delta^{\Gamma_+ Y}_{xz\mathbf{k}} \label{BCS3.1} \\
\Delta_\mathbf{k}^{\Gamma_-} &=& (v^{\Gamma}_{\mathbf{k}})^2 \Delta^{\Gamma_- X}_{yz\mathbf{k}} + (u^{\Gamma}_{\mathbf{k}})^2 \Delta^{\Gamma_- Y}_{xz\mathbf{k}}  \label{BCS3.2}\\
\Delta_\mathbf{k}^{X} &=& (u^{X}_{\mathbf{k}})^2 ( \Delta^{X\Gamma_+ }_{yz\mathbf{k}} + \Delta^{X\Gamma_-}_{yz\mathbf{k}} ) \label{BCS3.3} \\
\Delta_\mathbf{k}^{Y} &=& (u^{Y}_{\mathbf{k}})^2 ( \Delta^{Y \Gamma_+}_{xz\mathbf{k}} + \Delta^{Y\Gamma_ -}_{xz\mathbf{k}})  \label{BCS3.4}
\end{eqnarray}
where each low-energy band gap involves the sum of the different orbital contributions weighted by the correspondent coherent factors of the pocket. 
The gap functions $\Delta_\mathbf{k}^{l}$ given by the set of coupled equations in \ref{BCS3.1}\,-\,\ref{BCS3.4}, contain information on the spatial and orbital structures of the pairs.  The full set of equations, including also the contributions coming from the $xy$-pairing channel can be found in App.~\ref{App.C}.  

We solve numerically the linearized gap equations by searching for the largest eigenvalue that corresponds to the leading instability of the system. Then, we calculate its corresponding eigenfunction which  determines the symmetry and the structure of the gap function. We find an anisotropic $s_\pm$ symmetry as the dominant solution, followed by a nearly degenerate $d_{x^2-y^2}$ state in agreement with previous multiorbital RPA results \cite{Graser2009, ReviewHirschfeld, GapsymmetryHirschfeld2011, GapanisotropycomparitionRPA}. 

The results for the $s_\pm$ gap functions are summarized in Fig.s~\ref{fig:gapnodo}. Given the simplicity of our treatment, it is easy to verify that the overall momentum dependence of the band gaps directly follows from the momentum-dependence of the pairing vertex (i.e of the spin-susceptibility) and the orbital make-up of the Fermi surface. As already mentioned, the key parameter that controls the angular modulation of the gap is the orbital matching between the low-energy states connected by the spin-fluctuations that determine the structure of the spin-susceptibilities (and thus of the pairing vertex).
The orbital degree of freedom is also responsible for the hierarchy of the band gaps. In fact, given a similar condition of band nesting between the two hole pockets and the electrons, one could naively predict similar gaps opening along $\Gamma_\pm$. However, from Fig. \ref{fig:gapnodo}(b), it can be seen that the gap in  $\Gamma_-$ is larger than the gap in $\Gamma_+$. This is a direct consequence of the orbital composition of the nested pockets: in fact, the exchange of intraorbital spin-fluctuations between nested pockets having matching orbital, as $\Gamma_- - X/Y$ is stronger than the spin exchange between the pockets presenting orbital mismatch, such as $\Gamma_+ - X/Y$. This reflects in a larger spin-susceptibility and SC vertex for the pockets presenting a better degree of orbital matching. 

The results we obtain are in agreement with more complete RPA five-orbital model calculations. The main advantage of our derivation is that it gives us a new physical insight on the role of the orbital nesting in controlling the features of the SC gaps. We can compare our results, for example, with the ones obtained in \cite{GapanisotropycomparitionRPA}, where the anisotropy of the gaps for a five-orbital model is discussed in detail by using RPA calculations for the exchange of spin and charge fluctuations. The authors conclude that the anisotropy of the gap on the different Fermi surface pockets arises from an interplay of the orbital make-up of the states on the Fermi surface together with the momentum dependence of the fluctuation-exchange pairing interaction. In addition, to minimize the repulsion between the electron pockets, the gap can present accidental nodes. 
By comparing the results obtained within the OSSF model, Fig.s~\ref{fig:gapnodo}, and the five-orbital calculation, Fig. 4b in  \cite{GapanisotropycomparitionRPA}, one can see that the OSSF calculations qualitatively reproduce the main feature of the multiorbital ones i.e. the modulated $s_\pm$ symmetry for the gap, with accidental nodes. We also reproduce the correct hierarchy of the band gaps with $|\Delta_\mathbf{k}^{\Gamma_-}| > |\Delta_\mathbf{k}^{\Gamma_+}|$. The main novelty here is that our analysis allows for a straightforward interpretation of the gap anisotropy and the band gap hierarchy in terms of the degree of orbital matching between nested Fermi surfaces.  

%%%%%%%%%%%%%%%%%%%%%%%%%%%%%%%%%%%%%%%%%%
\section{Discussion}

In this work we show that the OSSF model provides a reliable description of the magnetic excitations in IBS and represents the minimal model to study spin-mediated superconductivity in this system. 

We compute the spin susceptibility for a four pocket model in the tetragonal phase and compare the results with the ones obtained for the five-orbital model. Depending on the degree of orbital nesting between pockets, we get an orbital modulation of the spin susceptibility that gives rise to commensurate or incommensurate peaks in the spin susceptibility when there is an orbital match or mismatch between the hole and the electron pockets respectively. The spin-exchange between pockets with matching orbitals is the larger contribution to the total spin-susceptibility.
By comparing the total spin susceptibility of the OSSF model with the one obtained within the five-orbital model we show that the OSSF reproduces qualitatively well the overall momentum dependence and the relative heights and widths of the peaks located at different momenta.
This is a remarkable result considering that the OSSF model is a low-energy effective model that only considers the $yz,xz,xy$ orbitals, and that the OSSF reduces the computation of the spin susceptibility to a few intraorbital scalar components.

We compute the pairing vertex mediated by OSSF and the corresponding gap equations. Due to the intraorbital scalar character of the spin susceptibility of the OSFF model it is possible to draw analytically the Feynman's diagrams involved in the pairing vertex, something which is almost unfeasible within the five-orbital model due to the large number of different possible diagrams. By solving the corresponding BCS equations we find as a leading instability a $s_\pm$ symmetry and a nearly degenerate  $d_{x^2-y^2}$ in agreement with multiorbital calculations. Our finding of a close competition between the $s_\pm$ ground state and the $d_{x^2-y^2}$ appears in agreement with Raman experiments in various IBS \cite{Loidl_PRL2013, HaiHu_PRX2014, Blumberg_PRB2017, Hackl_QM2018, 
Hack_PRB2018, Hackl_PRL2020}. The analysis of the subdominant $d_{x^2-y^2}$ symmetry within the OSSF model is beyond the scope of this work. However, we expect that the inclusion of the orbital degree of freedom within multiband approaches implemented so far (see e.g. \cite{Maiti_PRB2015,Maiti_PRL2016, Eremin_PRB2018, Eremin_PRB2019, Eremin_arxiv2021}) could provide further insight on the SC properties of IBS.

The analysis of the bands gap structure shows that the angular dependence and the magnitudes of the different gaps depend directly on the degree of orbital matching between the hole and the electron pockets. In particular, the orbital nesting between the different Fermi surfaces is the parameter that controls the hierarchy of the band gaps, in contrast with the naive expectation of having a gap proportional to the degree of band-nesting. 
This new insight on the role of the orbital composition of the nested pockets breaks the usual paradigm based on the Fermi surface matching as the parameter controlling the instabilities and establish the orbital nesting as the crucial control parameter. It is worth noticing that the connection between gap hierarchy and orbital matching of the nested pockets is more robust than a {\it naive} band-nesting argument as it does not depend on the fine tuning matching of size and shape of the Fermi surfaces, but relies instead on the orbital composition of the bands determined by the space group of the iron-plane of IBS.\\

In conclusion, in this work we analyze the magnetic excitations and SC gaps for a generic IBS system in the tetragonal phase using the OSSF model. Despite its simplicity, the OSSF model is able to reproduce well all the relevant qualitative features of the spin-excitations and pairing interactions of the multiorbital description. It allows for analytical treatment making the interpretation of the results straightforward. Due to the orbital-selectivity of the spin-fluctuation, it allows to include in a very simple way additional interacting channels, as we showed explicitly by considering the electron-electron interaction besides the hole-electron spin-exchange. The simplified frame of the OSSF allows us to gain physical insight on the role of orbital nesting in determining the angular modulations and the hierarchy of the band gaps. 
This result proves that the OSSF model is the minimal low-energy model to study spin-mediated superconductivity in IBS as it correctly incorporates in the low-energy description the orbital composition of the bands determined by the space group of the iron-plane.  

We thank A.~Kreisel and I.~Eremin for  useful  discussions. L.~F. acknowledges financial support from the European Unions Horizon 2020 research and innovation programme under the Marie Sk\l odowska-Curie grant SuperCoop (Grant No 838526). We acknowledge funding from Ministerio de Ciencia e Innovación, Agencia Estatal de Investigación and FEDER funds (EU) via Grant No. FIS2015-64654-P and  PGC2018-099199-B-I00.

\appendix 

\section{Kinetic Hamiltonian for the four-pocket model} 
\label{App.A}

The kinetic Hamiltonian considered in this work, given in Eq.~\ref{h0}, is adapted from the low-energy model of \cite{Vafekmodelo}. It consists in a four-pocket model with two hole-pockets at $\G$, $\G_\pm$ and two electron-pockets at $X$ and $Y$. Each pocket is described using a spinor representation in the pseudo-orbital space $H_0^l=\sum_{\bk,\s} 
\psi^{\dagger l}_{\bk\s} \hat H_{0 \bk}^l \psi^l_{\bk\s}$, where $l=\G,X,Y$ and the spinors are defined as: $\psi^{\Gamma}_{\bk 
\s}=(c^{yz}_{\bk,\s},c^{xz}_{\bk \s})$ and $\psi^{X/Y}_{\bk \s}=(c^{yz/xz}_{\bk \s},c^{xy}_{\bk \s})$. The matrix $\hat H_{0 \bk}^l $ has the general form
\be
\lb{a}
\hat H_{0 \bk}^l=
h_0^l\hat{\t_0}+\vec{h}^l \cdot \vec{\hat{\t}} = 
\begin{pmatrix}
h_0^l + h_3^l \ \ & \ \ h_1^l - ih_2^l \\
 h_1^l + ih_2^l \ \ & \ \ h_0^l - h_3^l \\
\end{pmatrix}
\ee
with $\hat{\t}$ matrices representing the pseudo-orbital spin. The $h^\G$ components read as
\bea
h_0^{\G} &=& \epsilon^\G -a^\G \bk^2, \nn \\
h_1^{\G} &=& -2 b^\G k_x k_y, \nn \\
h_3^{\G} &=& b^\G (k_x^2 -k_y^2),
\lb{hG}
\eea
and for the $X$ pocket,
\bea
h_0^{X} &=& (h^{yz}+ h^{xy})/2 \nn \\
h_2^{X} &=& v k_y \nn\\
h_3^{X} &=& (h^{yz}- h^{xy})/2 - b (k_x^2 -k_y^2) 
\lb{hX}
\eea
where $h^{yz} = -\epsilon^{yz} + a^{yz} \bk^2$ and 
$h^{xy}= - \epsilon^{xy} + a^{xy} \bk^2$. Analogous expressions hold for the $Y$ pocket provided that one exchange $k_x$ by $k_y$.   
Diagonalizing $\hat H^l_0$ we find the dispersion relations and the orbital composition for the bands $
H_0^l=\sum_{\mathbf{k},\sigma} 
\phi^{\dagger l}_{\mathbf{k}\sigma} \hat \Lambda_{0 \mathbf{k}}^l \phi^l_{\mathbf{k}\sigma}.
\label{Vafek_H}$
with $\phi^l = \hat {\mathcal U}^l \psi^{l}$ the fermionic operator in the band basis and $\hat {\Lambda}^l = \hat {\mathcal U}^l  \hat H_{0}^l \hat {\mathcal U}^{l^{-1}}= diag(E^{l_+},E^{l_-})$ the diagonal matrix containing the band dispersions $ E^{l_\pm} = h_{0}^l \pm h^l$ with $h^l = |\vec{h}^l|$. The components of the unitary matrix $\hat {\mathcal U}^l$, that connect the orbital-space to the band-space, are the coherence factors that represent the orbital content of the $l_{\pm}$-pockets. All the above quantities still depends on momentum and spin, we drop those labels to make the equations more readable.
Notice that in order to lift the degeneracy of the inner and outer $xz/yz$ pockets at $\Gamma$ we need to account for the spin-orbit coupling in the Hamiltonian. We added it explicitly by replacing $h^\Gamma\rightarrow \sqrt{(h^\Gamma)^2 + \lambda^2/4}$ in the expression for $E^{\Gamma^\pm}$.

\section{RPA Spin-Suceptibility for the OSSF model} 
\label{App.B}

The generic multiorbital spin susceptibility is a four-index tensor obtained from the analytical continuation  $i\Omega_{m} \rightarrow \Omega + i0^{+}$ of the Matsubara spin-spin correlation function  
\begin{equation}
\chi^{\eta \etap}_{\delta \deltap}(\mathbf{q}, i\Omega_{m}) = \int_{0}^{\beta} d\tau \, e^{i \Omega_{m} \tau} \,  \bigg\langle T_{\tau}  \vec{S}^{\eta\delta} (\mathbf{q}, \tau) \vec{S}^{\etap \deltap} (\mathbf{-q}, 0)  \bigg\rangle .
\label{App.RPA.1}
\end{equation}
where $\vec{S}^{\eta\delta} (\mathbf{q}, \tau)=\sum_{\bk ss'} \, c^{\eta \dagger}_{\bk s} \vec \s _{s s'}c^\delta_{\bk+\bq s'}$ is the spin operator in the orbital space. Substituting $\vec{S}^{\eta\delta} (\mathbf{q}, \tau)$ and applying the Wick's theorem to Eq. \ref{App.RPA.1} the spin susceptibility can be rewritten as 
\begin{equation} 
\chi^{\eta \etap}_{\delta \deltap}(\mathbf{q}, i\Omega_m) = - \frac{1}{\beta} \sum_{\mathbf{k},n} Tr \left[G^{\delta \eta}  (\mathbf{k},i\omega_n)
 G^{\deltap \etap} (\mathbf{k}+\mathbf{q},i\omega_n+i\Omega_m)  \right] 
\label{App.RPA.2}
\end{equation}
where the Green's function $G^{\delta \eta} $ is given by the rotation to the orbital space of the non-interacting Green function in the band basis. 

Within the OSSF model Eq. \ref{App.RPA.1} is significantly simplified. Due to the orbital-selective nature of the spin fluctuations, the spin operator becomes intraorbital $\vec{S}^\eta (\bq)=\sum_{\bk ss'} \, c^{\eta \dagger}_{\bk s} \vec \s _{s s'}c^\eta_{\bk+\bq s'}$, thus the spin-susceptibility tensor already reduce to a $2 \times 2$ matrix. The low-energy projection further simplifies the calculation reducing the analysis of the spin-susceptibility to the calculation of a few scalar components. Thus, Eq. \ref{App.RPA.2} within the OSSF model reads
\begin{equation} 
\chi^{l l'}(\mathbf{q}, i\Omega_m) = - \frac{1}{\beta} \sum_{\mathbf{k},n} Tr \left[\hat{G}^l(\mathbf{k},i\omega_n)
 \hat G^{l'} (\mathbf{k}+\mathbf{q},i\omega_n+i\Omega_m)  \right] 
\label{App.RPA.3}
\end{equation} 
The Green's functions of the OSSF model are defined around the high symmetry points $\Gamma, X, Y$ only and can be written in terms of the rotation matrices $\hat{\cal U}^l$ that diagonalize the $l$-Hamiltonian and of the Green's functions in the band basis $g^{l_\pm}_{(\mathbf{k},i\omega_{n})}$ as
\begin{align}
\begin{split}
\hat{G}^l (\mathbf{k},i\omega_n) \quad &= \quad {\hat{\cal U}^l} (\mathbf{k},i\omega_n) \bigg(diag \big(g^{l_{+}}(\mathbf{k},i\omega_{n}),g^{l_{-}}(\mathbf{k},i\omega_{n})\big) \bigg)^{-1}{\hat{\cal U}^{l^{-1}}}(\mathbf{k},i\omega_n) \quad = \\ \quad &= \quad 
\begin{pmatrix}
u^l & -v^l \\
v^{*l} & u^{*l}\\
\end{pmatrix}  
\begin{pmatrix}
g^{l_+} & 0 \\
0 & g^{l_-}\\
\end{pmatrix}
\begin{pmatrix}
u^{*l} & v^{l} \\
-v^{*l} & u^{l}\\
\end{pmatrix} 
\end{split}
\label{App.RPA.4}
\end{align}
By using these definitions into Eq. \ref{App.RPA.3} and performing the trace, the spin susceptibility associated to the spin exchange between the  $l_\pm,l^{'}_\pm$ pockets reads
\begin{equation}
\chi^{l_\pm l'_\pm}(\mathbf{q}, i\Omega_m) = - \frac{1}{\beta} \sum_{\mathbf{k},n} g^{l_\pm} (\mathbf{k},i\omega_n) g^{l'_\pm} (\mathbf{k}+\mathbf{q},i\omega_n+i\Omega_m) |(u/v)^{l}_{\mathbf{k}}|^2 |(u/v)^{l'}_{\mathbf{k}+\mathbf{q}}|^2
\label{App.RPA.5}
\end{equation}
Performing the Matsubara frequency summation and setting $\Omega \rightarrow 0$, we find the static susceptibility as given in Eq.\ref{SFOSSF3} of the main text. 
The complete set of orbital selective spin-susceptibilities of the model is given by
\begin{eqnarray}
\chi^{\Gamma_+ X}(\mathbf{q}) = - \sum_{\mathbf{k}} \frac{f(E^{X}_{\mathbf{k}+\mathbf{q}})-f(E^{\Gamma_+}_{\mathbf{k}})}{E^{X}_{\mathbf{k}+\mathbf{q}}-E^{\Gamma_+}_{\mathbf{k}}} |u^{\Gamma}_{\mathbf{k}}|^2 |u^{X}_{\mathbf{k}+\mathbf{q}}|^2 \nonumber \\
\chi^{\Gamma_- X}(\mathbf{q}) = - \sum_{\mathbf{k}} \frac{f(E^{X}_{\mathbf{k}+\mathbf{q}})-f(E^{\Gamma_-}_{\mathbf{k}})}{E^{X}_{\mathbf{k}+\mathbf{q}}-E^{\Gamma_-}_{\mathbf{k}}}|v^{\Gamma}_{\mathbf{k}}|^2 |u^{X}_{\mathbf{k}+\mathbf{q}}|^2 \nonumber \\
\chi^{\Gamma_+ Y}(\mathbf{q}) = - \sum_{\mathbf{k}} \frac{f(E^{Y}_{\mathbf{k}+\mathbf{q}})-f(E^{\Gamma_+}_{\mathbf{k}})}{E^{Y}_{\mathbf{k}+\mathbf{q}}-E^{\Gamma_+}_{\mathbf{k}}} |v^{\Gamma}_{\mathbf{k}}|^2 |u^{Y}_{\mathbf{k}+\mathbf{q}}|^2 \nonumber \\
\chi^{\Gamma_- Y}(\mathbf{q}) = - \sum_{\mathbf{k}} \frac{f(E^{Y}_{\mathbf{k}+\mathbf{q}})-f(E^{\Gamma_-}_{\mathbf{k}})}{E^{Y}_{\mathbf{k}+\mathbf{q}}-E^{\Gamma_-}_{\mathbf{k}}}|u^{\Gamma}_{\mathbf{k}}|^2 |u^{Y}_{\mathbf{k}+\mathbf{q}}|^2 \nonumber \\
\chi^{X Y}(\mathbf{q}) = - \sum_{\mathbf{k}} \frac{f(\epsilon^{Y}_{\mathbf{k}+\mathbf{q}})-f(\epsilon^{X}_{\mathbf{k}})}{\epsilon^{Y}_{\mathbf{k}+\mathbf{q}}-\epsilon^{X}_{\mathbf{k}}}|v^{X}_{\mathbf{k}}|^2 |v^{Y}_{\mathbf{k}+\mathbf{q}}|^2 .
\label{App.RPA.7}
\end{eqnarray}
Here we include the most relevant spin excitations between $\Gamma_{\pm}$ and $X/Y$ centered at $\mathbf Q_X=(\pi,0)$ and $\mathbf Q_Y=(0,\pi)$ momentum and having $yz$ and $xz$ orbital character respectively, as well as the spin susceptibility around $\mathbf Q_M=(\pi,\pi)$ resulting from the spin exchange between the $X-Y$ pockets and having $xy$ orbital character. 

The RPA spin susceptibilities are obtained in the form of Dyson-type equations as 
\begin{eqnarray}
\chi^{\Gamma_+ X}_{RPA}(\mathbf{q})  &=& \frac{\chi^{\Gamma_+ X}(\mathbf{q})}{1- \tilde{U}\chi^{\Gamma_+ X}(\mathbf{q})} \qquad \chi^{\Gamma_+ Y}_{RPA}(\mathbf{q})  = \frac{\chi^{\Gamma_+ Y}(\mathbf{q})}{1- \tilde{U}\chi^{\Gamma_+ Y}(\mathbf{q})} \qquad \chi^{X Y}_{RPA}(\mathbf{q})  = \frac{\chi^{X Y}(\mathbf{q})}{1- \tilde{U}\chi^{X Y}(\mathbf{q})}  \nonumber \\
\chi^{\Gamma_- X}_{RPA}(\mathbf{q})  &=& \frac{\chi^{\Gamma_- X}(\mathbf{q})}{1- \tilde{U}\chi^{\Gamma_- X}(\mathbf{q})} \qquad \chi^{\Gamma_- Y}_{RPA}(\mathbf{q})  = \frac{\chi^{\Gamma_- Y}(\mathbf{q})}{1- \tilde{U}\chi^{\Gamma_- Y}(\mathbf{q})} 
\label{App.RPA.9}
\end{eqnarray}
where $\tilde{U}$ is the intraorbital effective coupling and $\chi^{l_\pm l'_\pm}(\mathbf{q})$ are the ones given in Eq. \ref{App.RPA.7}. The RPA suscpetibilities given in Eqs. \ref{App.RPA.9} are those represented in Fig.s \ref{fig:122RPA} and \ref{fig:CaminoGXMG}a of the main text. The model parameters used in the numerical evaluation are reported in App.\ref{App.C}. 

\section{BCS gap equations} 
\label{App.C}

We consider the BCS Hamiltonian for the $yz/xz$ orbital sector Eq.~\ref{BCS1} of the main text and the $xy$-pairing term resulting from the pair hopping between the $X-Y$ electron pockets and explicitly given by
\begin{eqnarray}
H^{pairing}_{xy} = - \sum_{\mathbf{k},\mathbf{k'}} \Gamma^{X Y}_{xy \ \mathbf{k}\mathbf{k'}} \left[(v^X_{\mathbf{k}})^2 e^{X}_{\mathbf{k'}} e^{X}_{\mathbf{-k'}}  (v^Y_{\mathbf{k}})^2 e^{Y}_{\mathbf{k'}} e^{Y}_{\mathbf{-k'}}  \right] + h.c. 
\label{BCS1xy}
\end{eqnarray}

The mean-field equations for the total pairing Hamiltonian, Eqs.~\ref{BCS1} and ~\ref{BCS1xy}, can be easily derived by defining the orbital-dependent superconducting order parameters for the hole sector ($\Delta_{yz}^{\Gamma_+X},\Delta_{yz}^{\Gamma_-X},\Delta_{xz}^{\Gamma_+Y},\Delta_{xz}^{\Gamma_-Y}$) and the electron sector ($\Delta_{yz}^{X\Gamma_+},\Delta_{yz}^{X\Gamma_-},\Delta_{xy}^{XY},\Delta_{xz}^{Y\Gamma_+},\Delta_{xz}^{Y\Gamma_-},\Delta_{xy}^{YX})$) as
\begin{eqnarray}
\Delta^{\Gamma_+X}_{yz \mathbf{k'}}&=& - \Gamma_{yz \mathbf{k}\mathbf{k'}}^{\Gamma_+X} \langle (u^{X}_{\mathbf{k}})^{2} e^{X}_{\mathbf{k}} e^{X}_{\mathbf{-k}} \rangle   \nn \\
\Delta^{\Gamma_-X}_{yz \mathbf{k'}}&=& - \Gamma_{yz \mathbf{k}\mathbf{k'}}^{\Gamma_-X} \langle (u^{X}_{\mathbf{k}})^{2} e^{X}{\mathbf{k}} e^{X}_{\mathbf{-k}} \rangle   \nn \\
\Delta^{YX}_{xy \mathbf{k'}}&=& - \Gamma^{X Y}_{xy \ \mathbf{k}\mathbf{k'}} \langle (v^{X}_{\mathbf{k}})^{2} e^{X}_{\mathbf{k}} e^{X}_{\mathbf{-k}} \rangle   \nn \\
\Delta^{\Gamma_+Y}_{xz \mathbf{k'}}&=& - \Gamma_{xz\mathbf{k}\mathbf{k'}}^{\Gamma_+Y} \langle (u^{Y}_{\mathbf{k}})^{2} e^{Y}{\mathbf{k}} e^{Y}_{\mathbf{-k}} \rangle   \nn \\
\Delta^{\Gamma_-Y}_{xz \mathbf{k'}}&=& - \Gamma_{xz\mathbf{k}\mathbf{k'}}^{\Gamma_-Y} \langle (u^{Y}_{\mathbf{k}})^{2} e^{Y}_{\mathbf{k}} e^{Y}_{\mathbf{-k}} \rangle   \nn \\
\Delta^{X\Gamma_+}_{yz \mathbf{k'}}&=& - \Gamma_{yz\mathbf{k}\mathbf{k'}}^{\Gamma_+X} \langle (u^{* \Gamma}_{\mathbf{k}})^{2} h^{+}_{\mathbf{k}} h^{+}_{\mathbf{-k}} \rangle   \nn \\
\Delta^{X\Gamma_-}_{yz \mathbf{k'}}&=& - \Gamma_{yz\mathbf{k}\mathbf{k'}}^{\Gamma_-X} \langle (v^{\Gamma}_{\mathbf{k}})^{2} h^{-}_{\mathbf{k}} h^{-}_{\mathbf{-k}} \rangle   \nn \\
\Delta^{Y\Gamma_+}_{xz \mathbf{k'}}&=& - \Gamma_{xz\mathbf{k}\mathbf{k'}}^{\Gamma_+Y} \langle (v^{* \Gamma}_{\mathbf{k}})^{2} h^{+}_{\mathbf{k}} h^{+}_{\mathbf{-k}} \rangle   \nn \\
\Delta^{Y\Gamma_-}_{xz \mathbf{k'}}&=& - \Gamma_{xz\mathbf{k}\mathbf{k'}}^{\Gamma_-Y} \langle (u^{ \Gamma}_{\mathbf{k}})^{2} h^{-}_{\mathbf{k}} h^{-}_{\mathbf{-k}} \rangle   \nn \\
\Delta^{XY}_{xy \mathbf{k'}}&=& - \Gamma^{X Y}_{xy \ \mathbf{k}\mathbf{k'}} \langle (v^{Y}_{\mathbf{k}})^{2} e^{Y}{\mathbf{k}} e^{Y}_{\mathbf{-k}} \rangle   \nn \\
\label{App.BSC.1}
\end{eqnarray}
The corresponding self-consistent BCS equations at $T=0$ are  
\begin{eqnarray}
\Delta^{\Gamma_+X}_{yz\mathbf{k'}}&=&-\sum_{k} \frac{\Gamma^{\Gamma_+ X}_{yz\mathbf{k}\mathbf{k'}} }{|v^X_{F\mathbf{k}}|} (u^{X}_{\mathbf{k}})^{2} \left[ (u^{X}_{\mathbf{k}})^{2} \Delta^{X\Gamma_+}_{yz\mathbf{k}} + (u^{X}_{\mathbf{k}})^{2} \Delta^{X\Gamma_-}_{yz\mathbf{k}} + (v^{X}_{\mathbf{k}})^{2} \Delta^{XY}_{xy\mathbf{k}} \right]  \label{BCSxy2.1}  \\
\Delta^{\Gamma_-X}_{yz\mathbf{k'}} &=& - \sum_{k} \frac{\Gamma^{\Gamma_- X}_{yz\mathbf{k}\mathbf{k'}} }{|v^X_{F\mathbf{k}}|} (u^{X}_{\mathbf{k}})^{2} \left[ (u^{X}_{\mathbf{k}})^{2} \Delta^{X\Gamma_+}_{yz\mathbf{k}} + (u^{X}_{\mathbf{k}})^{2} \Delta^{X\Gamma_-}_{yz\mathbf{k}} + (v^{X}_{\mathbf{k}})^{2} \Delta^{XY}_{xy\mathbf{k}} \right]  \label{BCSxy2.2}  \\
\Delta^{YX}_{xy\mathbf{k'}} &=& - \sum_{k} \frac{\Gamma^{X Y}_{xy \mathbf{k}\mathbf{k'}}}{|v^X_{F\mathbf{k}}|} (v^{X}_{\mathbf{k}})^{2} \left[ (u^{X}_{\mathbf{k}})^{2} \Delta^{X\Gamma_+}_{yz\mathbf{k}} + (u^{X}_{\mathbf{k}})^{2} \Delta^{X\Gamma_-}_{yz\mathbf{k}} + (v^{X}_{\mathbf{k}})^{2} \Delta^{XY}_{xy\mathbf{k}} \right] 
\label{BCSxy2.10} \\
\Delta^{\Gamma_+Y}_{xz\mathbf{k'}} &=& - \sum_{k} \frac{\Gamma^{\Gamma_+ Y}_{xz\mathbf{k}\mathbf{k'}} }{|v^Y_{F\mathbf{k}}|} (u^{Y}_{\mathbf{k}})^{2} \left[ (u^{Y}_{\mathbf{k}})^{2}  \Delta^{Y\Gamma_+}_{xz\mathbf{k}} + (u^{Y}_{\mathbf{k}})^{2} \Delta^{Y\Gamma_+}_{xz\mathbf{k}} + (v^{Y}_{\mathbf{k}})^{2} \Delta^{YX}_{xy\mathbf{k}} \right]  \label{BCSxy2.3}  \\
\Delta^{\Gamma_-Y}_{xz\mathbf{k'}} &=&  - \sum_{k} \frac{\Gamma^{\Gamma_- Y}_{xz\mathbf{k}\mathbf{k'}} }{|v^Y_{F\mathbf{k}}|} (u^{Y}_{\mathbf{k}})^{2}  \left[(u^{Y}_{\mathbf{k}})^{2}  \Delta^{Y\Gamma_+}_{xz\mathbf{k}} + (u^{Y}_{\mathbf{k}})^{2}  \Delta^{Y\Gamma_-}_{xz\mathbf{k}} + (v^{Y}_{\mathbf{k}})^{2} \Delta^{YX}_{xy\mathbf{k}} \right]  \label{BCSxy2.4} \\
\Delta^{XY}_{xy\mathbf{k'}} &=& - \sum_{k} \frac{\Gamma^{X Y}_{xy \mathbf{k}\mathbf{k'}}} {|v^Y_{F\mathbf{k}}|} (v^{Y}_{\mathbf{k}})^{2}  \left[(u^{Y}_{\mathbf{k}})^{2}  \Delta^{Y\Gamma_+}_{xz\mathbf{k}} + (u^{Y}_{\mathbf{k}})^{2}  \Delta^{Y\Gamma_-}_{xz\mathbf{k}} + (v^{Y}_{\mathbf{k}})^{2} \Delta^{YX}_{xy\mathbf{k}} \right] 
\label{BCSxy2.9} %\\
\label{BCS2}
\end{eqnarray}
plus Eqs.~\ref{BCS2.5}-\ref{BCS2.8} of the main text that remain the same once the $xy$-pairing is included in the analysis. This set of coupled BCS equations is solved numerically using band parameters given in App.\ref{App.C}. The band gaps reported in Fig.\ref{fig:gapnodo} are the gap functions defined in terms of the orbital-dependent order parameters as 
\begin{eqnarray}
\Delta_\mathbf{k}^{\Gamma_+} &=& (u^{\Gamma}_{\mathbf{k}})^2 \Delta^{\Gamma_+X}_{yz\mathbf{k}} + (v^{\Gamma}_{\mathbf{k}})^2 \Delta^{\Gamma_+Y}_{xz\mathbf{k}} \label{BCSxy3.1} \\
\Delta_\mathbf{k}^{\Gamma_-} &=& (v^{\Gamma}_{\mathbf{k}})^2 \Delta^{\Gamma_-X}_{yz\mathbf{k}} + (u^{\Gamma}_{\mathbf{k}})^2 \Delta^{\Gamma_-Y}_{xz\mathbf{k}}  \label{BCSxy3.2}\\
\Delta_\mathbf{k}^{X} &=& (u^{X}_{\mathbf{k}})^2 \Delta^{X\Gamma_+}_{yz\mathbf{k}} + (u^{X}_{\mathbf{k}})^2 \Delta^{X\Gamma_-}_{yz\mathbf{k}} + (v^{X}_{\mathbf{k}})^{2} \Delta^{XY}_{xy\mathbf{k}}  \label{BCSxy3.3} \\
\Delta_\mathbf{k}^{Y} &=& (u^{Y}_{\mathbf{k}})^2 \Delta^{Y\Gamma_+}_{xz\mathbf{k}} + (u^{Y}_{\mathbf{k}})^2 \Delta^{Y\Gamma_-}_{xz\mathbf{k}} + (v^{Y}_{\mathbf{k}})^{2} \Delta^{YX}_{xy\mathbf{k}} \label{BCSxy3.4}
\end{eqnarray}

\section{Band parameters used in the calculations} 
\label{App.D}

In the calculations of the RPA spin-susceptibility shown in Fig.s \ref{fig:122RPA} and \ref{fig:CaminoGXMG}a we use for the kinetic Hamiltonian the set of parameters given in Table \ref{Table.Kinetic}, and fix $\lambda$ to $5$ meV. Those parameters are the ones that reproduce the four-pocket model shown in Fig. \ref{fs} of the main text.

\begin{table}[tbh]
\begin{center}
\begin{tabular}{ccccccccccccccccccc}
\hline 
&                           &$\Gamma$&                     &\qquad \qquad \qquad&         &X/Y&                                          & \\
\hline \hline
&$\epsilon_\Gamma$ &           & 46&                    &$\epsilon_{xy}$ \ \      72        & &$\epsilon_{yz/xz}$\ \  55 &  \\
\hline
& $a_\Gamma$          &           &263&                   & $a_{xy}$          \ \      93        & &$a_{yz/xz}$\ \           101&  \\
\hline
& $b_\Gamma$          &           &182&                   & $b$                  \ \ \   154      & &$v$ \ \                                        144&  \\
\hline
\hline
\end{tabular}
\caption{Model parameters for a generic four-pockets system. All the parameters are in meV.} 
\vspace{-0.5cm}
\label{Table.Kinetic}
\end{center}
\end{table} 

In the analysis of the band gaps Fig.\ref{fig:gapnodo} we use a slightly different band structure with hole-pockets having similar size, i.e. the same degree of band-nesting with the electron pockets, to better emphasize the effect of the orbital composition of the nested Fermi surface. In order to that we fix $\epsilon_\Gamma = 10$ meV, $a_\Gamma=150$ meV and $b_\Gamma=50$ meV. The electron bands parameters are instead the same of Table\ref{Table.Kinetic}

\bibliography{pnictides_nematic}
\end{document}